\documentclass[doublecol]{epl2}
\usepackage{amsfonts}
\usepackage{amssymb}
\usepackage{amsmath}

\nocite{*}
\institute{
\inst{1} Department of Physics, Tohoku University, Sendai, 980-8578, Japan\\
\inst{2} Condensed Matter Theory Group, CPMOH, UMR 5798, Universit\'{e} Bordeaux I,33405 Talence, France
} 
\abstract{A graphene $pn$ junction is studied theoretically
in the presence of both intrinsic and Rashba spin-orbit couplings. We show that a crossover from perfect reflection to perfect transmission 
is achieved at normal incidence by tuning the perpendicular electric field.
By further studying angular dependent transmission, we demonstrate that
perfect reflection at normal incidence can be clearly distinguished 
from trivial band gap effects. We also investigate how spin-orbit effects modify the conductance 
and the Fano factor associated with a potential step in both $nn$ and $np$ cases.}
\input{tcilatex}

\begin{document}

\title{Spin-orbit effects in a graphene bipolar pn junction}
\author{A. YAMAKAGE$^{1}$, K.-I. IMURA$^{1,2}$, J. CAYSSOL$^{2}$ and Y.
KURAMOTO$^{1}$}
\maketitle

\section{Introduction}

There has been recent interest in a novel class of band insulators, called
topological insulators (TIs) \cite{revue}. TIs are characterized by a bulk
gap and a pair of time-reversed edge states at the boundary. These gapless
edge states originate from the lattice spin-orbit (SO) effect and are
protected by time-reversal symmetry from moderate disorder and interaction.
In their seminal paper \cite{kane05}, Kane and Mele demonstrated that a
graphene monolayer may become a TI when the intrinsic SO coupling dominates
over the extrinsic Rashba coupling. Then the graphene layer is characterized
by a $Z_{2}$ topological number \cite{kane05b} and has gapless edge states
that disappear if the Rashba coupling becomes too large, or if Coulomb
interactions exceed a certain threshold \cite{wu06}. Therefore the existence
of the topological phase in graphene depends crucially on the values of the
intrinsic and extrinsic (Rashba) SO couplings. In the concluding section, we
shall discuss recent estimates of these SO interactions which allows for
considering the transition between the topological ($\Delta \geq \lambda
_{R} $) and ordinary ($\lambda _{R}\geq \Delta $) phases.

Since the insulating regime is rather difficult to access experimentally, we
propose to characterize this crossover by transport properties in the doped
regime. In particular we stress that transport through a bipolar $pn$
junction should differ in the distinct cases $\Delta \geq \lambda _{R}$ and $%
\lambda _{R}\geq \Delta $ respectively. Quasi-relativistic Klein tunneling 
\cite{klein29} was demonstrated experimentally \cite%
{huard07,williams07,ozyilmaz07,oostinga07,gorbatchev08,stander09,liu08} by
using local gating techniques, and the corresponding theory have received a
great deal of attention \cite{katsnelson06,cheianov06,cayssol09,sonin09} in
the absence of SO coupling ($\Delta =\lambda _{R}=0$).

In this Letter, we obtain the angular dependent transmission of a $pn$
junction in the presence of both intrinsic and Rashba SO couplings. At
normal incidence, we show that the $pn$ junction transmission exhibits a
crossover from perfect reflection to perfect transmission when the Rashba
coupling is tuned by the perpendicular electric field. We also predict the
conductance and the Fano factor associated with a potential step in presence
of intrinsic and Rashba SO couplings.

Similar topological phases have also been predicted \cite{bernevig06} and
now observed \cite{konig07} in materials with larger SO interactions such as
HgTe/CdTe quantum wells. The protected edge states appear when the width of
the HgTe layer exceeds a critical value. We choose to study graphene since
the Kane-Mele is the simplest possible model that have four spin-split
bands, which is the minimum required for the nontrivial phase to exist \cite%
{murakami07,hughes08}. Moreover our predictions might be compared to the
studies of Klein tunneling in graphene performed in the absence of SO
coupling ($\Delta =\lambda _{R}=0$).

\section{Kane-Mele model and single valley approximation}

The Kane-Mele model describes the low-energy dynamics of quasiparticles near
the $K$ and $K^{\prime }$ points of graphene in the presence of spin-orbit
effects \cite{kane05}. The corresponding Hamiltonian $%
H_{KM}=H_{0}+H_{SO}+H_{R}$ acts on the slowly varying envelop $\psi (x,y)$
of electronic Bloch wavefunctions, which are indexed by real spin (Pauli
matrices $s_{i}$, $i=x,y,z$), lattice isospin ($\sigma _{i}$) and valley
isospin ($\tau _{i}$) quantum numbers. The kinetic Hamiltonian 
\begin{equation}
H_{0}=-i\hbar v_{F}\psi ^{\dagger }(\sigma _{x}\tau _{z}\partial _{x}+\sigma
_{y}\partial _{y})\psi
\end{equation}%
describes massless Dirac fermions and is spin-independent. In the following
we shall use units with $\hbar =v_{F}=1$.

The intrinsic spin-orbit effect is completely determined by the symmetries
of the honeycomb lattice and by the geometry of the carbon orbitals. It can
be described by the Hamiltonian 
\begin{equation}
H_{SO}=-\Delta \psi ^{\dagger }\sigma _{z}\tau _{z}s_{z}\psi ,
\end{equation}%
where $2\Delta $ is the value of the gap induced at $K$ (and $K^{\prime }$).
This form can be deduced from group theoretical techniques \cite{kane05} or
as the low-energy limit of tight-binding models \cite{min06,huertas06,yao07}%
. It was shown recently that $\Delta $ can be increased by inducing a
curvature of the graphene layer \cite{huertas06}.

In the presence of a perpendicular electric field (generated by the distant
gate), there is an additional Rashba spin-orbit coupling \cite{bychkov84}

\begin{equation}
H_{R}=\lambda _{R}\psi ^{\dagger }(\sigma _{y}s_{x}-\sigma _{x}\tau
_{z}s_{y})\psi ,  \label{HKM}
\end{equation}%
where $\lambda _{R}$ is proportional to the electric field.

In the following we shall restrict ourselves to transport through potential
barriers which are smooth on the scale of the atomic lattice period.
Therefore we shall neglect intervalley scattering by using a single-valley
version of the Kane-Mele Hamiltonian $H_{KM}$ wherein the eigenvalue of $%
\tau _{z}$ is replaced by $+1$(or $-1$ ) for $K$ (or $K^{\prime }$). The
resulting Hamiltonian $H_{KM}^{(K)}$ consists in a $4\times 4$ matrix
applying to spinors of the form $^{t}\left[ \psi _{A\uparrow },\psi
_{B\uparrow },\psi _{A\downarrow },\psi _{B\downarrow }\right] $, where $%
^{t} $ represents transpose. Here the arrow index ($\uparrow $, $\downarrow $%
) stands for real spin while the index ($A,B$) denotes the two inequivalent
sites of the honeycomb lattice. 
\begin{figure*}[tbp]
\begin{minipage}{0.33\hsize}
\includegraphics{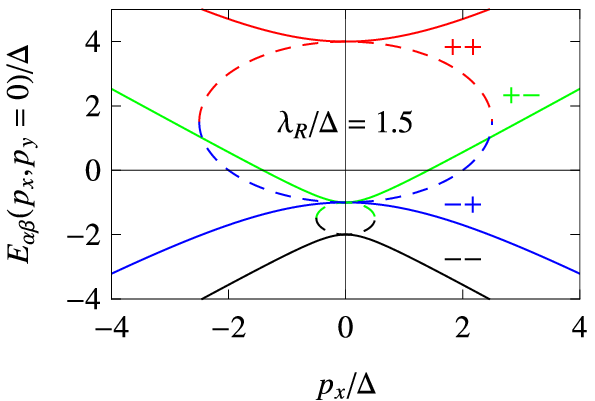}
\end{minipage}
\begin{minipage}{0.33\hsize}
\begin{center}
\includegraphics{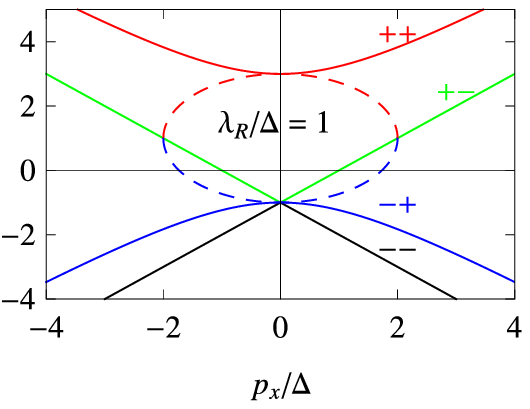}
\end{center}
\end{minipage}
\begin{minipage}{0.33\hsize}
\begin{center}
\includegraphics{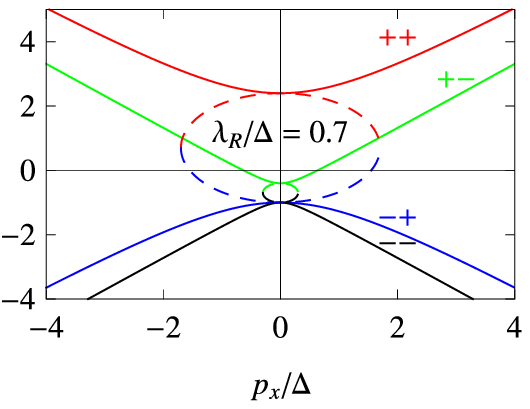}
\end{center}
\end{minipage}
\caption{Energy spectra for different values of $\protect\lambda _{R}/\Delta
=$ 1.5, 1, 0.7. Propagating modes are shown in solid curves while evanescent
modes are superposed in broken curves with $p_{x}$ replaced by $\protect%
\kappa $ in Eq. (\protect\ref{kappa}). When Rashba coupling dominates the
intrinsic SO interaction $\Delta $ (left panel), the spectrum is gapless and
quadratic, whereas it is gapped in the opposite case (right panel). In the
balanced case $\protect\lambda _{R}=\Delta $ (central panel) the spectrum is
gapless and hosts a Dirac cone. }
\label{spectrum}
\end{figure*}

\section{Band structure}

In the homogeneous case, the two-dimensional momentum $\mathbf{p}%
=(p_{x},p_{y})$ is a good quantum number. The single valley Kane-Mele
Hamiltonian $H_{KM}^{(K)}$ is diagonalized by the eigenspinors $|\alpha
\beta \rangle _{\mathbf{p}}$ as 
\begin{equation}
H_{KM}^{(K)}|\alpha \beta \rangle _{\mathbf{p}}=E_{\alpha \beta }(\mathbf{p}%
)|\alpha \beta \rangle _{\mathbf{p}},
\end{equation}%
where\ $\alpha ,\beta =\pm 1$ are band indices. The four energy bands are
characterized by the dispersion 
\begin{equation}
E_{\alpha \beta }(\mathbf{p})=\alpha \sqrt{\mathbf{p}^{2}+(\Delta +\beta
\lambda _{R})^{2}}+\beta \lambda _{R},  \label{Eab}
\end{equation}%
where $\mathbf{p}^{2}\mathbf{=}p_{x}^{2}+p_{y}^{2}$. The energy spectra has
distinct features depending on the value of the ratio $\lambda _{R}/\Delta $
(Fig. \ref{spectrum}).

In the absence of Rashba coupling, the intrinsic spin-orbit coupling opens a
gap at $\mathbf{p=0}$ separating the conduction band $E_{+\beta }(\mathbf{p}%
)=\sqrt{\mathbf{p}^{2}+\Delta ^{2}}$ from the valence band $E_{-\beta }(%
\mathbf{p})=-\sqrt{\mathbf{p}^{2}+\Delta ^{2}}$, each one still having a
two-fold degeneracy with respect to $\beta =\pm 1$.

A finite Rashba interaction generates a splitting of those bands which may
eventually close the gap. In that sense the intrinsic and Rashba SO effects
compete each other. When the intrinsic SO dominates ($\Delta >\lambda _{R}$%
), the system is a band insulator whereas it is a gapless semimetal with
quadratically dispersing bands in the opposite case ($\Delta <\lambda _{R}$%
). The intermediate case $\Delta =\lambda _{R}$ is rather special since two
linearly dispersing bands $E_{\alpha -}(\mathbf{p})=\alpha \left\vert 
\mathbf{p}\right\vert -\lambda _{R}$ are recovered. Nevertheless the system
differs from the graphene spectrum in the absence of SO coupling $E(\mathbf{p%
})=\pm v_{F}\left\vert \mathbf{p}\right\vert $ by the presence of two
additional spin-split parabolic bands $E_{\alpha +}(\mathbf{p})$.

In coordinate representation, the eigenspinors $|\alpha \beta \rangle _{%
\mathbf{p}}$ are plane waves $\left\langle \mathbf{r}|\alpha \beta
\right\rangle _{\mathbf{p}}=\Phi _{\beta }(E,\mathbf{p})e^{i\mathbf{p.r}}$
with 
\begin{equation}
\Phi _{\beta }(E,\mathbf{p})=A(E,\mathbf{p})\left[ 
\begin{array}{c}
p_{x}-ip_{y} \\ 
E+\Delta \\ 
-i\beta \left( E+\Delta \right) \\ 
-i\beta (p_{x}+ip_{y})%
\end{array}%
\right] ,  \label{spinor}
\end{equation}%
and $\mathbf{p.r=}p_{x}x+p_{y}y$. The normalization factor is simply 
\begin{equation*}
A(E,\mathbf{p})={\frac{1/\sqrt{2}}{\sqrt{|p_{x}|^{2}+p_{y}^{2}+\left( E(%
\mathbf{p})+\Delta \right) ^{2}}}.}
\end{equation*}%
Note that for a given energy $E$, the band index $\alpha $ is automatically
determined. The spinor Eq. (\ref{spinor}) also describes evanescent modes
characterized by a purely imaginary $p_{x}$ (superposed in broken curves in
Fig. \ref{spectrum}). Naturally, those modes cannot exist in an infinite
sheet, but they develop in the presence of an interface (see below the
interface between the $n-$ and $p-$ doped regions) and contribute to the
scattering properties.

Eigenstates with different momentum or energy are obviously orthogonal to
each other ${}_{\mathbf{p}}\langle \alpha \beta |\alpha ^{\prime }\beta
^{\prime }\rangle _{\mathbf{p}^{\prime }}=\delta _{\alpha \alpha ^{\prime
}}\delta _{\beta \beta ^{\prime }}\delta _{\mathbf{pp}^{\prime }}$. Here we
amphasize that the eigenstates Eq. (\ref{spinor}) of the Kane-Mele model
satisfy an additional orthogonality relation: $\Phi _{\beta }^{\dag }\Phi
_{\beta ^{\prime }}=\delta _{\beta \beta ^{\prime }}$ when $p_{y}=0$.
Moreover the band indices $\alpha $ and $\beta $ play very different roles:
index $\beta $ determines the symmetry of wave function, while $\alpha $
specifies whether the state belongs to the conduction or the valence band.
In Fig. \ref{spectrum} (b), two energy bands with the same band index $\beta
=-$ combine to form a linear Dirac spectrum. The orthogonality between
eigenstates with different $\beta $ also holds true for evanescent modes
realized in the vicinity of $pn$-junction.

\section{The $pn$-junction Model}

We now introduce our model of a $pn$-junction in a monolayer of graphene
described within the Kane-Mele model. We assume that an electrostatic gate
creates a potential barrier $V(x)$ which is smooth on the scale of the
atomic lattice. Moreover we consider an idealy pure system, a situation
which have been approached experimentally in suspended devices \cite%
{andrei08,bolotin08}. Then no inter-valley scattering is involved, i.e. $K$
and $K^{\prime }$ points are decoupled. One can use safely the single-valley
approximation and describe the junction by the $4$-band Hamiltonian%
\begin{equation}
H(x)=H_{KM}^{(K)}+V(x)\psi ^{\dagger }\psi ,  \label{scattering}
\end{equation}%
where the $V(x)$ term is diagonal in both spin and lattice isospin degrees
of freedom. Moreover we assume that the barrier is sharp on the scale of the
Fermi wavelength in each of the metallic bands at right and left. In this
sense, we shall investigate the scattering by an abrupt barrier defined by 
\begin{equation}
V(x)=\left\{ 
\begin{array}{ll}
0 & (x<0) \\ 
V_{0}>0 & (x>0)%
\end{array}%
\right. .  \label{step}
\end{equation}%
Note that we also assume a straight interface with translational invariance
along the $y$ direction (no roughness along the interface $x=0$).

\section{Scattering states}

We investigate how incident fermions from the left are scattered by the
potential barrier $V(x)$. The left side ($x<0$) being always in the doped
metallic regime, transport across the junction is dominated by extended
two-dimensional waves while helical edge states are irrelevant here. Owing
to translational invariance along the $y$ axis, the momentum $p_{y}$ is a
good quantum number and factors $e^{ip_{y}y}$ can be omitted accordingly. We
now construct scattering states at the Fermi level defined by their energy $E
$ and momentum $p_{y}$. We choose the energy $E$ such that incident
particles are injected from the single band $E_{+-}(\mathbf{p})$, which is
realized for $-\Delta <E<2\lambda _{R}+\Delta $.

The scattering state on the incident side takes the following form:

\begin{equation}
\Psi (x<0)=\Phi ^{(i)}e^{ikx}+r\Phi ^{(r)}e^{-ikx}+r_{ev}\Phi
^{(ev)}e^{\kappa x},
\end{equation}%
where $\Phi ^{(i)}=\Phi _{-}(E,\mathbf{p}_{i})$, $\Phi ^{(r)}=\Phi _{-}(E,%
\mathbf{p}_{r})$, and $\Phi ^{(ev)}=\Phi _{+}(E,\mathbf{p}_{ev})$. The
wavevectors of the incident, reflected and evanescent waves are respectively 
$\mathbf{p}_{i}=(k,p_{y}),\mathbf{p}_{r}=(-k,p_{y})$ and $\mathbf{p}%
_{ev}=(-i\kappa ,p_{y})$, with 
\begin{eqnarray}
k &=&\sqrt{(E+\Delta )(E-\Delta +2\lambda _{R})-p_{y}^{2}},  \label{k} \\
\kappa &=&\sqrt{p_{y}^{2}-(E+\Delta )(E-\Delta -2\lambda _{R})}.
\label{kappa}
\end{eqnarray}

On the transmitted side, the wavefunction reads%
\begin{equation}
\Psi (x>0)=t_{+}\Phi ^{(+)}e^{ip_{+}x}+t_{-}\Phi ^{(-)}e^{ip_{-}x},
\end{equation}%
where $\Phi ^{(\beta )}=\Phi _{\beta }(E-V_{0},\mathbf{p}_{\beta })$ and $%
\mathbf{p}_{\beta }=(p_{x\beta },p_{y})$ are amplitude and wavevector of the
two transmitted waves $\beta =\pm $. The spinors $\Phi ^{(+)}$ and $\Phi
^{(-)}$ represent either a propagating or an evanescent ($p_{\pm }$ becomes
pure imaginary) mode depending on the sign of 
\begin{equation}
p_{x\beta }^{2}=(E-V_{0}+\Delta )(E-V_{0}-\Delta -2\beta \lambda
_{R})-p_{y}^{2}.  \label{pbeta}
\end{equation}%
If $p_{x\beta }^{2}>0$, the mode in the corresponding $E_{-\beta }(\mathbf{p}%
)$ band is propagating. The actual sign of $p_{x\beta }$ is chosen such that
the group velocity is positive, thereby describing an outgoing transmitted
wave packet. In the specific case of inter-band tunneling, the positive
group velocity is realized by a negative momentum state ($\alpha <0$ in Eq. (%
\ref{Eab})) implying $p_{x\beta }<0$. If $p_{x\beta }^{2}<0$, the mode in
the corresponding $E_{-\beta }(\mathbf{p})$ band is evanescent.

Demanding continuity of the wavefunctions $\Psi (0^{+})=\Psi (0^{-})$ at the
interface: 
\begin{equation}
\Phi ^{(i)}+r\Phi ^{(r)}+r_{ev}\Phi ^{(ev)}=t_{+}\Phi ^{(+)}+t_{-}\Phi
^{(-)},  \label{cc}
\end{equation}%
we obtain four independent scalar equations for the scattering parameters : $%
r$
, $r_{ev}$, $t_{+}$ and $t_{-}$. Thus the reflection probability $R=|r|^{2}$
is determined uniquely from Eq. (\ref{cc}) for given $E$, $V_{0}$ and $p_{y}$%
.

\section{Transmission at normal incidence}

We examine here the normal incidence transmission ($p_{y}=0$) through the $%
pn $ junction in the presence of Rashba and intrinsic SO effects. The $pn$
junction is defined by Eq. (\ref{step}) together with the condition $%
V_{0}-E>\Delta $ to insure inter-band tunneling.

\subsection{Decoupling at normal incidence}

When $p_{y}=0$, the continuity equation Eq. (\ref{cc}) reduces to two 
decoupled equations 
\begin{eqnarray}
r_{ev}\Phi ^{(ev)} &=&t_{+}\Phi ^{(+)},  \label{cc+} \\
\Phi ^{(i)}+r\Phi ^{(r)} &=&t_{-}\Phi ^{(-)},  \label{cc-}
\end{eqnarray}%
due to the symmetry of the spinors. Indeed, the full Hilbert space is the
direct sum $\Omega =\Omega _{+}\oplus \Omega _{-}$ of two orthogonal
subspaces. The spinors $\Phi ^{(ev)}$and $\Phi ^{(+)}$ belong to the
subspace $\Omega _{+}$ spanned by $^{t}\left[ 1,0,0,-i\right] $ and $^{t}%
\left[ 0,1,-i,0\right] $, where $^{t}$ represents transpose. The spinors $%
\Phi ^{(i)},\Phi ^{(r)}$ and $\Phi ^{(-)}$ belong to the orthogonal subspace 
$\Omega _{-}$ spanned by $^{t}\left[ 1,0,0,i\right] $ and $^{t}\left[ 0,1,i,0%
\right] $.

Therefore solving Eq.(\ref{cc-}) yields the following simple expression for
the reflection amplitude at normal incidence 
\begin{equation}
r=\frac{k(E-V_{0}+\Delta )-p_{x-}(E+\Delta )}{k(E-V_{0}+\Delta
)+p_{x-}(E+\Delta )},  \label{rnormal}
\end{equation}%
where $k$ and $p_{x-}$ are obtained by substituing $p_{y}=0$ in Eqs.(\ref{k},%
\ref{pbeta}).

\subsection{Dominant Rashba effect}

At large enough Rashba coupling, namely $\lambda _{R}>(\Delta +V_{0}-E)/2,$
there is one propagating ($\Phi ^{(+)}$) and one evanescent ($\Phi ^{(-)}$)
transmitted waves, momenta $p_{x+}$ and $p_{x-}$ being respectively real and
purely imaginary. As a result, Eq.(\ref{rnormal}) yields unimodular
reflection amplitude $r$, thereby indicating perfect reflection. This
nontrivial total reflection arises because the only propagating transmitted
wave ($\Phi ^{(+)}$) is orthogonal to the incident wave ($\Phi ^{(i)}$) at
normal incidence. This situation is similar to inter-band tunneling in
bilayer graphene as we shall discuss later. For smaller Rashba coupling $%
\Delta <\lambda _{R}<(\Delta +V_{0}-E)/2,$ both transmitted waves are
propagating leading to finite transmission through the $\Phi ^{(-)}$ wave.
Accordingly\ the momentum $p_{x-}$ is real and Eq.(\ref{rnormal}) implies
partial reflection, i.e. $0<\left\vert r\right\vert ^{2}<1$.

\begin{figure}[tbp]
\par
\begin{center}
\includegraphics[width=7cm]{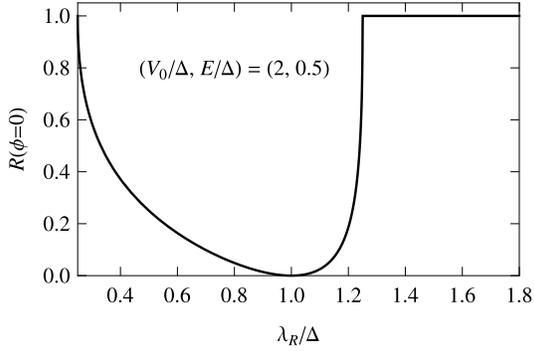}
\end{center}
\par
\caption{Reflection probability $R(\protect\phi =0)$ of the $pn$ junction as
a function of $\ \protect\lambda _{R}/\Delta $, at normal incidence.}
\label{normal}
\end{figure}

\subsection{Balanced Rashba and intrinsic SO effects}

At $\lambda _{R}=\Delta $, the bands $E_{\alpha -}$ become linear $E_{\alpha
-}(\mathbf{p})=\alpha \left\vert \mathbf{p}\right\vert -\Delta $ and combine
to form a Dirac cone. Meanwhile, the spinors show further orthogonality
relations in addition to the one with respect to $\beta $. Namely, at this
particular value the reflected wave becomes orthogonal to the incident wave,
thereby implying perfect transmission. In the case of (mono-layer) graphene
in the absence of SO effects, the total absence of backscattering at normal
incidence is due to the chiral symmetry, i.e. $T^{2}=-1$ ($T$: time-reversal
operator), and Berry phase $\pi $ \cite{ando98, katsnelson06}. Another
important consequence of chiral symmetry is the anti-localization in the
absence of inter-valley scattering \cite{suzuura02}. Here, in the Kane-Mele
model, $T^{2}=1$ and the Berry phase is $2\pi $, due to the activation \cite%
{imura09} of real spin by Rashba SO coupling. As a result the system shows
standard weak localization in the absence of inter-valley scattering \cite%
{imura09}.

\subsection{Topological gap phase}

At large intrinsic SO coupling $\lambda _{R}<\Delta $, both $\Phi ^{(+)}$
and $\Phi ^{(-)}$ describe propagating waves. Therefore the reflection is
only partial.

Summarizing this section, we have seen that the $pn$ junction shows a
crossover from perfect reflection at large Rashba coupling towards perfect
transmission when $\lambda _{R}=\Delta $, while finite reflection is
restored at smaller values of the Rashba coupling (Fig. \ref{normal}). These
contrasted behaviors are reminiscent of those of a $pn$ junction in single
and bilayer graphene which show respectively perfect transmission and
perfect reflection at normal incidence \cite{katsnelson06,guinea09}. These
remarkable features originate from the orthogonality between the incident
and scattered spinors at normal incidence. However, this orthogonality
relation between $\Phi _{+}(E,\mathbf{p})$ and $\Phi _{-}(E,\mathbf{p})$ is
broken as soon as $p_{y}$ becomes finite.

It is worthy to note that the parameter $i\lambda _{R}$ in the Kane-Mele
model plays formally the role of inter-layer hopping in bilayer graphene.
Thus monolayer graphene with only Rashba SO has the same structure as that
of spinless bilayer graphene, and shows the same charge transport properties.

\begin{figure}[tbp]
\begin{center}
\includegraphics[width=8.5cm]{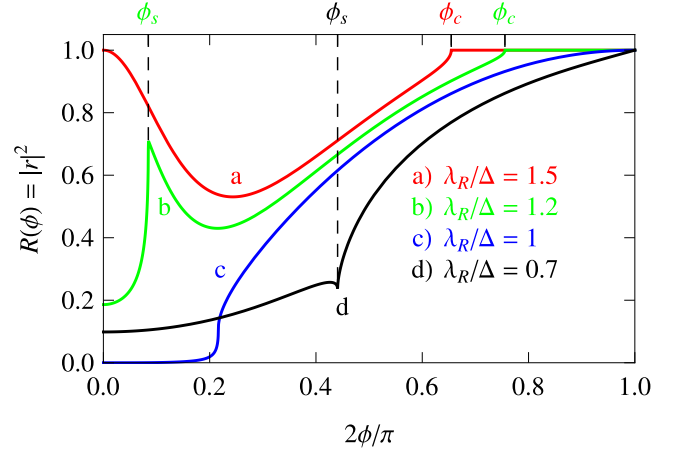}
\end{center}
\caption{Angular dependence of the reflection probability $R(\protect\phi %
)=|r|^{2}$. $R(\protect\phi )$ is plotted at $(V_{0}/\Delta ,E_{0}/\Delta
)=(2,0.5)$ for different values of $\protect\lambda _{R}/\Delta =1.5$, $1.2$%
, $1$, $0.7$. Four different types of behaviors can be seen: (a) crossover
from perfect reflection to partial transmission with a broad dip ($\protect%
\lambda _{R}/\Delta =1.5$), (b) a peak in $R(\protect\phi )$ ($\protect%
\lambda _{R}/\Delta =1.2$), (c) crossover from perfect to partial
transmission ($\protect\lambda _{R}/\Delta =1$), and (d) a cusp like dip in $%
R(\protect\phi )$ ($\protect\lambda _{R}/\Delta =0.7$).}
\label{angular}
\end{figure}

\section{Transmission at arbitrary incidence}

We now focus on the angular dependence of the transmission through a bipolar 
$pn$ junction. The reflection probability $R(\phi )=\left\vert r\right\vert
^{2}$ is obtained by solving the continuity equation Eq.(\ref{cc}), the
incident angle $\phi $ satisfying $p_{y}=k\sin \phi $.

\subsection{Dominant Rashba effect ($\protect\lambda _{R}>\Delta )$}

\begin{figure}[tbp]
\par
\begin{center}
\includegraphics[scale=0.65]{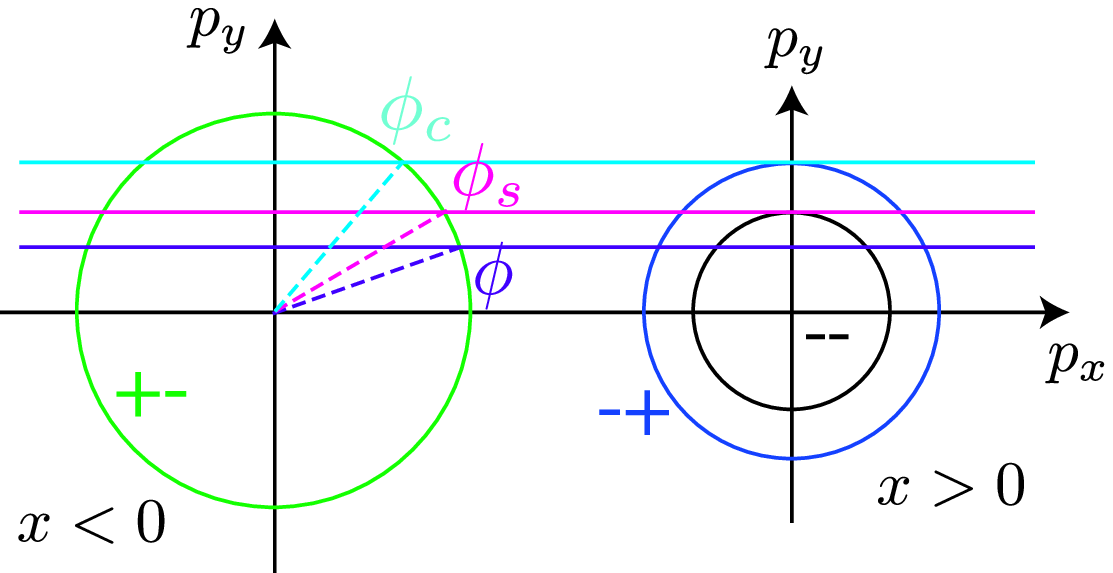}
\end{center}
\par
\par
\begin{center}
\includegraphics[scale=0.55]{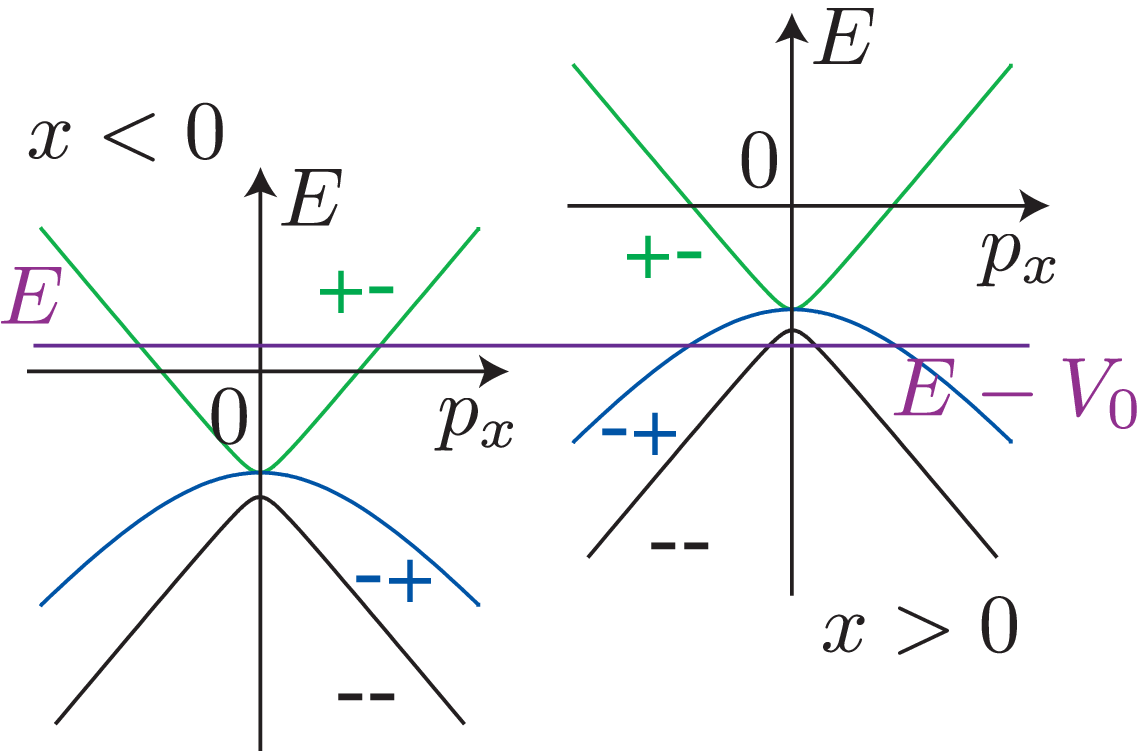}
\end{center}
\par
\caption{Fermi circles (upper panel) and energy bands (lower panel) for
intermediate Rashba coupling: $\Delta <\protect\lambda _{R}<(\Delta
+V_{0}-E)/2$ on both sides of the junction. When incident angle $\protect%
\phi $ is smaller than a critical value $\protect\phi _{s}$, both of the
transmitted states are propagating. At $\protect\phi =\protect\phi _{s}$, $%
\Phi ^{(-)}$ mode turns to evanescent. When $\protect\phi $ exceeds $\protect%
\phi _{c}$, both of the transmitted states become evanescent. A pair of $+$
or $-$ indices refer to the band indices $\protect\alpha \protect\beta $.}
\label{circles}
\end{figure}

As discussed previously the $pn$ junction exhibits a perfect reflection at
normal incidence for large enough Rashba coupling $\lambda _{R}>(\Delta
+V_{0}-E)/2$. As one varies the incident angle $\phi $ from normal incidence
($\phi =0$), the reflection coefficient decreases and the curve $R(\phi )$
exhibits a broad dip (curve $\lambda _{R}/\Delta =1.5$, Fig. \ref{angular}%
.a). This feature clearly distinguishes the perfect reflection due to
orthogonality in the normal incidence from the perfect reflection due to a
band gap on the transmitted side. In the case of perfect reflection due to
band gap, the reflection probability remains trivially equal to unity when
one varies the incident angle $\phi $ away from normal incidence. At large
incidence $\phi >\phi _{c}$, one recovers total reflection. The critical
angle $\phi _{c}$ is determined by a condition on the Fermi wavevectors in
the bands $E_{+-}$ and $E_{-+}$ as shown in Fig. \ref{circles}.

For intermediate Rashba coupling $\Delta <\lambda _{R}<(\Delta +V_{0}-E)/2$,
reflection is only partial at $\phi =0$ and the curve $R(\phi )$ exhibits a
peak (curve $\lambda _{R}/\Delta =1.2$, Fig. \ref{angular}.b). The initial
increase of $R(\phi )$ from $\phi =0$ to $\phi _{s}$ is related to the
overlap between the incident wave and transmitted ($\Phi ^{(+)}$ and $\Phi
^{(-)}$) waves. At small incidence, the dominant effect is the reduction of
the overlap with the $\Phi ^{(-)}$ mode yielding an increasing reflection
probability. The local maximum of $R(\phi )$ at $\phi =\phi _{s}$ appears
when the $\Phi ^{(-)}$ mode turns to evanescent (Fig. \ref{circles}, upper
panel). For slightly larger incidence $\phi >\phi _{s}$, the dominant effect
is the increase of the overlap between the incident wave and the
(propagating) transmitted $\Phi ^{(+)}$ mode, thereby providing a decrease
of $R(\phi )$. Above the critical angle $\phi >\phi _{c}$, both transmitted
modes become evanescent at $x>0$, thereby leading to $R(\phi )=1$ (Fig. \ref%
{circles}).

\subsection{Balanced case $(\protect\lambda _{R}=\Delta )$}

When $\lambda _{R}=\Delta $, the property of perfect transmission (which is
exact at $\phi =0$) pertains quite accurately to a broad range around normal
incidence (curve $\lambda _{R}/\Delta =1$, Fig. \ref{angular}.c).

\subsection{Topological gap phase $(\protect\lambda _{R}<\Delta )$}

When $\lambda _{R}<\Delta $, the reflection probability $R(\phi )$ shows a
sharp dip (curve $\lambda _{R}/\Delta =1.2$, Fig. \ref{angular}.d) at the
angle $\phi _{s}$ where $\Phi ^{-}$ mode turns to evanescent (Fig. \ref%
{circles}, upper panel). The nature of this singularity is similar to that
of peak structure at intermediate Rashba coupling already discussed. The
singularity appears when $E-(\Delta -2\lambda _{R})>-\Delta -(E-V_{0})$,
which means $\lambda _{R}/\Delta >0.5$ for $(V_{0}/\Delta ,E_{0}/\Delta
)=(0.5,2)$. Note that $E_{+-}$ and $E_{--}$ bands are symmetric w.r.t. $%
E=-\lambda _{R}$.

\section{Conductance and shot noise}

\begin{figure*}[tbp]
\begin{minipage}{0.49\hsize}
\includegraphics{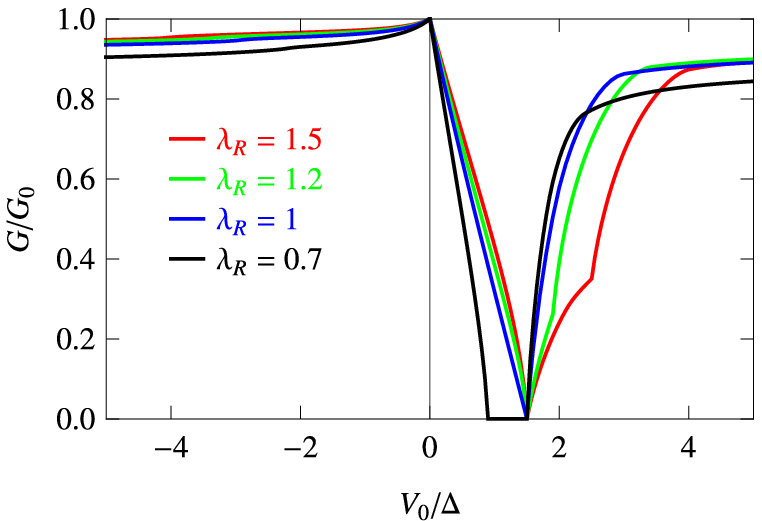}
\end{minipage}
\begin{minipage}{0.49\hsize}
\begin{center}
\includegraphics{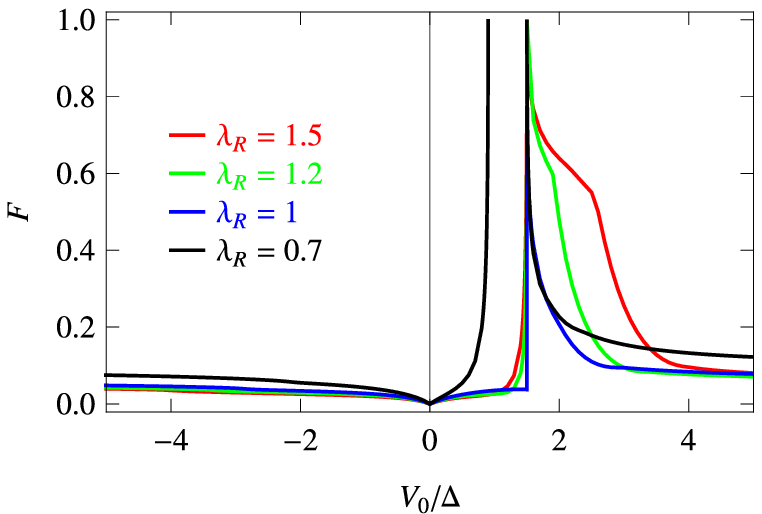}
\end{center}
\end{minipage}
\caption{Conductance $G$ (left) and Fano factor $F$ (right) of the junction
for $E=0.5\Delta $, and different values of $\protect\lambda _{R}/\Delta
=1.5,1.2,1,0.7$. The maximal conductance is given by $G_{0}=(2e^{2}/\protect%
\pi h)k_{F}W$ where $W$ is the sample width (along $y$), and $k_{F}$ is the
Fermi wavevector in the left side ($x<0$).}
\label{condfano1}
\end{figure*}

\textbf{\ }The conductance $G(V_{0})$ and the Fano factor $F(V_{0})$
associated with the potential step are readily obtained from the
transmission probability, as described in \cite%
{katsnelson06,cheianov06,cayssol09,sonin09}. In the balanced case $\lambda
_{R}=\Delta $, the conductance curves $G(V_{0})$ (Fig. 5, left panel) do not
differ significantly from the case of no SO effects ($\lambda _{R}=\Delta =0$%
) \cite{cayssol09,sonin09}. In contrast, the vanishing of the conductance
within a finite range of $V_{0}$ reveals that graphene is gapped when
intrinsic SO dominates Rashba coupling. The Fano factor is, on the other
hand, not well-defined in the gap. The Rashba dominated regime also exhibits
anomalous features: a cusp in $G(V_{0})$ and an enhancement of the Fano
factor. The peak in $F(V_{0})$ is rather asymmetric with a characteristic
shoulder that broadens when increasing the Rashba coupling (Fig. \ref%
{condfano1}, $\lambda _{R}/\Delta =1.2,1.5$).

The corresponding experiments should be done on graphene with very low
absolute Fermi energies on both sides of the junction. Therefore disorder
may hinder the observation of the crossover from perfect reflection to
perfect transmission at the $pn$-junction \cite{martin08}. Nevertheless
on-going progress in sample preparation might eventually render the
spin-orbit effects observable in suspended graphene devices \cite%
{andrei08,bolotin08}. The SO interaction is commonly supposed to be weak
owing to the low atomic number of carbon. Nevertheless, first estimates
indicated the favorable conditions for the TI to exist, namely a sizeable
spin-orbit gap $2\Delta \sim 1$ K at the $K,K^{\prime }$ points and a tiny
Rashba splitting $\lambda _{R}\sim 0.5$ mK for a typical electric field $%
E=50 $ V/$300$ nm \cite{kane05}. Unfortunately due to the specific geometry
of $s$ and $p$ orbitals in graphene, the actual intrinsic SO coupling should
be far smaller, namely $2\Delta \sim 10$ mK, while the Rashba splitting is
enhanced to typical values $\lambda _{R}\sim 70$ mK for $E=50$ V/$300$ nm 
\cite{min06,huertas06,yao07}. Recently, first-principle calculations
suggested that $d$ orbitals might play a dominant role in the gap opening at 
$K$ and $K^{\prime }$ points \cite{gmitra09}. As a result, the spin
couplings are predicted to assume intermediate values\ ($2\Delta \sim
\lambda _{R}\sim 0.1$ K) between the estimates of \cite{kane05} and of \cite%
{min06,huertas06,yao07}. Such close values and the possibility to tune both
Rashba and intrinsic SO couplings allows to consider the transition between
the topological ($\Delta \geq \lambda _{R}$) and ordinary ($\lambda _{R}\geq
\Delta $) phases.

\section{Concluding remarks}

We have shown that\ at normal incidence the $pn$ junction transmission
exhibits a crossover from perfect reflection at large Rashba coupling to
perfect transmission when the Rashba coupling exactly balances the intrinsic
spin orbit coupling. Further study on the angular dependence enabled us to
clearly distinguish such unique features from trivial band gap effects.
Finally we have obtained the conductance and the shot noise associated with
an electrostatic potential step realized in a graphene monolayer with
competing spin-orbit effects.

\acknowledgments A.Yamakage and K.I. Imura are supported by KAKENHI
(A.Yamakage: No. 08J56061 of MEXT, Japan, K.I. Imura: Grant-in-Aid for Young
Scientists B-19740189). J. Cayssol acknowledges gratefully support from the
Institut de Physique Fondamentale (IPF) in Bordeaux.

\end{document}